# Anapole-Assisted Strong Field Enhancement in Individual All-Dielectric Nanostructures

Yuanqing Yang,[*] Vladimir A. Zenin, and Sergey I. Bozhevolnyi

SDU Nano Optics, University of Southern Denmark, Campusvej 55, DK-5230 Odense, Denmark

*To whom correspondence should be addressed. Email address: yy@mci.sdu.dk.

**ABSTRACT**

High-index dielectric nanostructures have recently become prominent forefront alternatives for manipulating light at the nanoscale. Their electric and magnetic resonances with intriguing characteristics endow them with a unique ability to strongly enhance near-field effects with minimal absorption. Similar to their metallic counterparts, dielectric oligomers consisting of two or more coupled particles are generally employed to create localized optical fields. Here we show that *individual* all-dielectric nanostructures, with rational designs, can produce strong electric fields with intensity enhancements exceeding 3 orders of magnitude. Such a striking effect is demonstrated within a Si nanodisk by fully exploiting anapole generation and simultaneously introducing a slot area with high-contrast interfaces. By performing finite-difference time-domain simulations and multipole decomposition analysis, we systematically investigate both far-field and near-field properties of the slotted disk and reveal a subtle interplay among different resonant modes of the system. Furthermore, while electric fields at anapole modes are typically internal, i.e., found inside nanostructures, our slotted configuration generates *external* hotspots with electric fields additionally enhanced by virtue of boundary conditions. These electric hotspots are thereby directly accessible to nearby molecules or quantum emitters, opening up new possibilities for single-particle enhanced spectroscopies or single-photon emission enhancement due to large Purcell effects. Our presented design methodology is also readily extendable to other materials and other geometries, which may unlock enormous potential for sensing and quantum nanophotonic applications.

**INTRODUCTION**

Controlled creation of strong electromagnetic field enhancements is one of the most important and fundamental strategies to engineer light-matter interactions, giving rise to the modern realm of nanophotonics and propelling transformative breakthroughs in a diverse set of areas such as biosensing, energy harvesting, and information processing.[1] Metallic nanostructures featuring a multitude of localized and propagating surface-plasmon modes have received considerable attention over the past few decades due to their unprecedented ability to concentrate light into nanometric volumes.[2,3] However, high intrinsic absorption losses and associated local heating of plasmonic nanostructures severely limit their practical use in many scenarios. To circumvent this issue, high-index dielectric materials with low dissipative losses have recently been proposed as a promising alternative for manipulating light at the nanoscale.

Compared to their metallic counterparts whose resonances are often dominated by electric response, high-index dielectric nanoparticles can support a distinct series of Mie-type resonances with both electric and magnetic modes[4-10]. These excited dipole or multipole modes can further interfere with each other, thus providing new routes for engineering electromagnetic field distributions and obtaining strongly enhanced optical fields. In this context, coupled dielectric nanoresonators[11-24] are generally adopted to create localized hotspots, since near fields of single dielectric particles are inherently less confined and hence less enhanced. For example, dielectric nanodimers made of two cylinders[14] have been demonstrated to support both electric and magnetic hotspots under different polarizations, owing to their coupled electric and magnetic dipolar resonances. This geometry and rendered hotspots were later leveraged to produce fluorescence enhancements and surface-enhanced Raman scattering[16,17]. Other configurations, including oligomers[18-21] and arrays,[4-7, 22-24] have also been found to exhibit exotic responses such as Fano resonances and efficient nonlinear phenomena. Nevertheless, despite growing efforts to enhance near-field processes,[25-29] it remains a formidable challenge for all-dielectric nanostructures to attain high local fields that have comparable strengths to those of their plasmonic analogs, especially for the electric field components[15].

In this work, we show that strong electric fields can be obtained in *individual* all-dielectric nanostructures with intensity enhancements exceeding 3 orders of magnitude. Such a significant enhancement factor is achieved by utilizing optical anapole modes, where far-field scattering is minimized and near-field energy within dielectrics reaches its maximum.[30, 31] This tantalizing property of anapole modes makes dielectric nanoparticles virtually non-radiating sources with enhanced near fields, which can be of crucial importance to many applications such as cloaking,[32] harmonic generation,[33-35] and nanoscale lasers[36]. Despite that, the electric field enhancement at conventional anapoles is still relatively mild [30] ($|E_{max}/E_0|^2 \sim 10$) and the strongest electric hotspots are inside particles so that one cannot easily get access to them. Here, to overcome these limitations, we introduce a properly designed slot area at maxima of electric fields generated by anapole modes (see a representative slotted nanodisk in Figure 1). By doing this, the field enhancement at anapole modes can be further boosted via boundary conditions, to which the high permittivities of dielectric materials are also related and can contribute substantially. We also demonstrate that such a design does not bring any degradations in the far-field signatures of anapoles and can facilitate selective multi-spectral field enhancements caused by higher-order anapole modes. The geometry dependence of the effects is further systematically investigated. By resorting to multipole decomposition analysis, we unambiguously show the interplay among different resonant modes supported by the system. Moreover, the electric hotspots exposed by our slotted configuration become directly accessible to nearby emitters such as molecules, quantum dots or NV centers in diamond, which would be particularly beneficial for single-particle enhanced spectroscopies, sensing and quantum nanophotonic applications. On the basis of this notion, we finally demonstrate how our presented nanostructure can efficiently enhance single-photon emission with a large Purcell effect.

**RESULTS AND DISCUSSION**

Figure 1 shows a schematic illustration of a slotted Si disk under normally incident planewave illumination. To start with, we examine the scattering response of the system and consider a Si disk with $D$ = 600 nm and $H$ = 100 nm in free space as the first example. The length $L$ and the width $W$ of the slot are chosen as 200 nm and 10 nm, respectively. The polarization of the incoming beam in the $y$-direction is perpendicular to the long axis ($x$-direction) of the slot. Three-dimensional finite-difference time-domain (FDTD) simulations and total-field/scattered-field technique were employed to analyze the electromagnetic properties of the structure (see Methods). The dispersive optical constant of Si was taken from the experimental data determined by ellipsometric measurements of deposited amorphous Si (Figure S1).

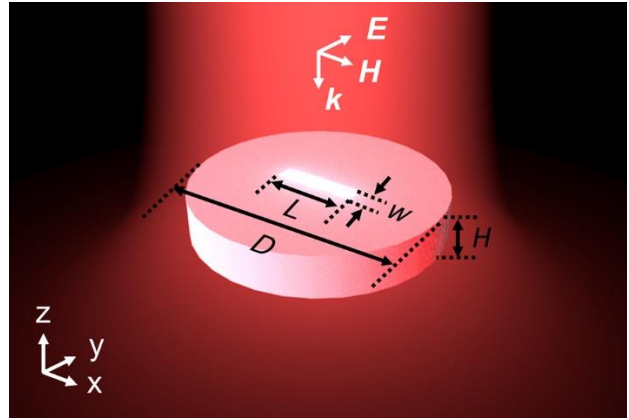

**Figure 1.** Schematic illustration of a slotted Si nanodisk under normal-incident illumination. The Si disk has a diameter of $D$ and a height of $H$. The slot in the center has a length of $L$ and a width of $W$. The polarization of the incident wave is along the *y*-axis, perpendicular to the long axis (*x* direction) of the slot. Light is confined inside the slot volume and thereby creates a bright hotspot in situ.

The scattering cross section of the slotted disk is shown in Figure 2a (solid black line). Two marked dips around 800 nm and 1200 nm can be clearly seen in the spectrum. To unambiguously reveal the excitation of both fundamental and higher-order anapole modes, we further decompose the scattering cross section of the structure into a series of spherical multipoles (see Supporting Information). The partial scattering from the spherical electric dipole (dotted blue line) exhibits two distinct minima that correspond well to the valleys of the total scattering, indicating the first- and second-order anapole modes, namely $AM_1$ and $AM_2$, excited in the system. Here by 'anapole modes', we refer to the scattering minima of the dominant electric dipole mode, following the conventional definition used in previous studies.[30, 31] Given the electric-magnetic duality, one may also expect anapole modes of magnetic character, as revealed and discussed in a recent work.[37]

Figure 2b provides the electric-field intensity enhancement calculated at the center of the disk. Two prominent peaks, with enhancement factors larger than 500 and 1000 respectively, coincide well with the positions of the two anapole modes shown in the scattering spectrum. A comprehensive comparison between the slotted disk and a solid disk is made in Figure S2. For the scattering cross section, the slotted disk manifests a slight blue shift compared to the solid one because of the reduced geometrical cross section and a shortened optical path inside the particle. Interestingly, although both disks exhibit distinctive anapole hallmarks in the far-field, the slotted disk shows a much more significant near-field enhancement compared to the solid disk.

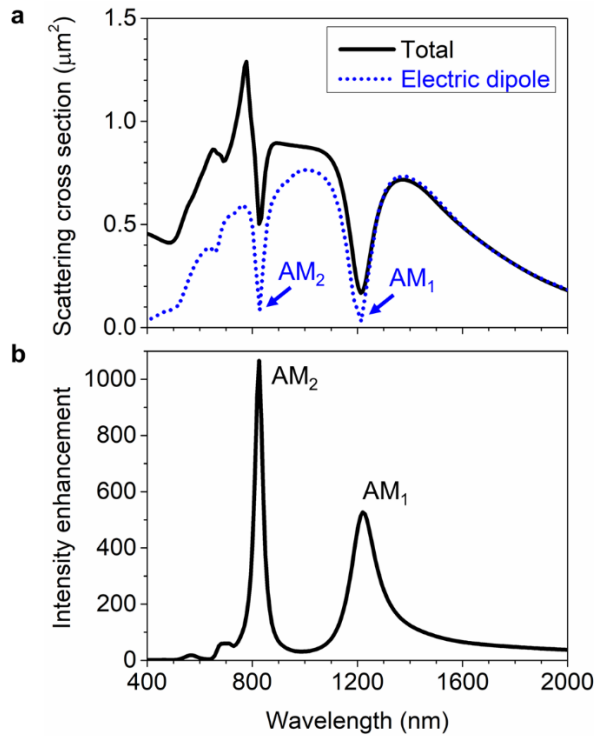

**Figure 2.** Far-field scattering and near-field enhancement of a representative slotted Si disk with $D = 600$ nm, $H = 100$ nm, $L = 200$ nm and $W = 10$ nm. (a) Scattering cross section of the slotted Si nanodisk. Total scattering spectrum (black solid line) and partial scattering contributed by the spherical electric dipole (blue dotted line) are depicted. (b) Intensity enhancement of the electric field at the center of the slotted disk.

To gain deeper physical insight into the giant field enhancement introduced by the slot, we plot near field distributions at the $AM_1$ and $AM_2$ modes, as displayed in Figure 3. Saturated fields are presented intentionally to reveal the anapole features of the slotted disk as now they can be directly related to the conventional anapoles of the solid disk (Figure S3). Similar to the solid disk, the $AM_1$ mode of the slotted disk shows two opposite circular displacement currents on the left and right half of the disk (Figure 3a). Two singular points with field magnitudes close to zero can be found along the $x$-axis (Figure 3b).[34] By contrast, the $AM_2$ mode supports two pairs of such poloidal currents and results in four field zeros in the $x$ direction (Figure 3d,e). The field distribution of the $AM_2$ mode indicates a clear combination of the $AM_1$ mode and an accompanied standing wave character, which can be explained by the formation of hybrid Mie-Fabry-Perot modes[38] or the superposition of several internal modes[39]. Moreover, for both solid and slotted disks, the $AM_2$ mode shows a larger field enhancement compared to the $AM_1$ mode.

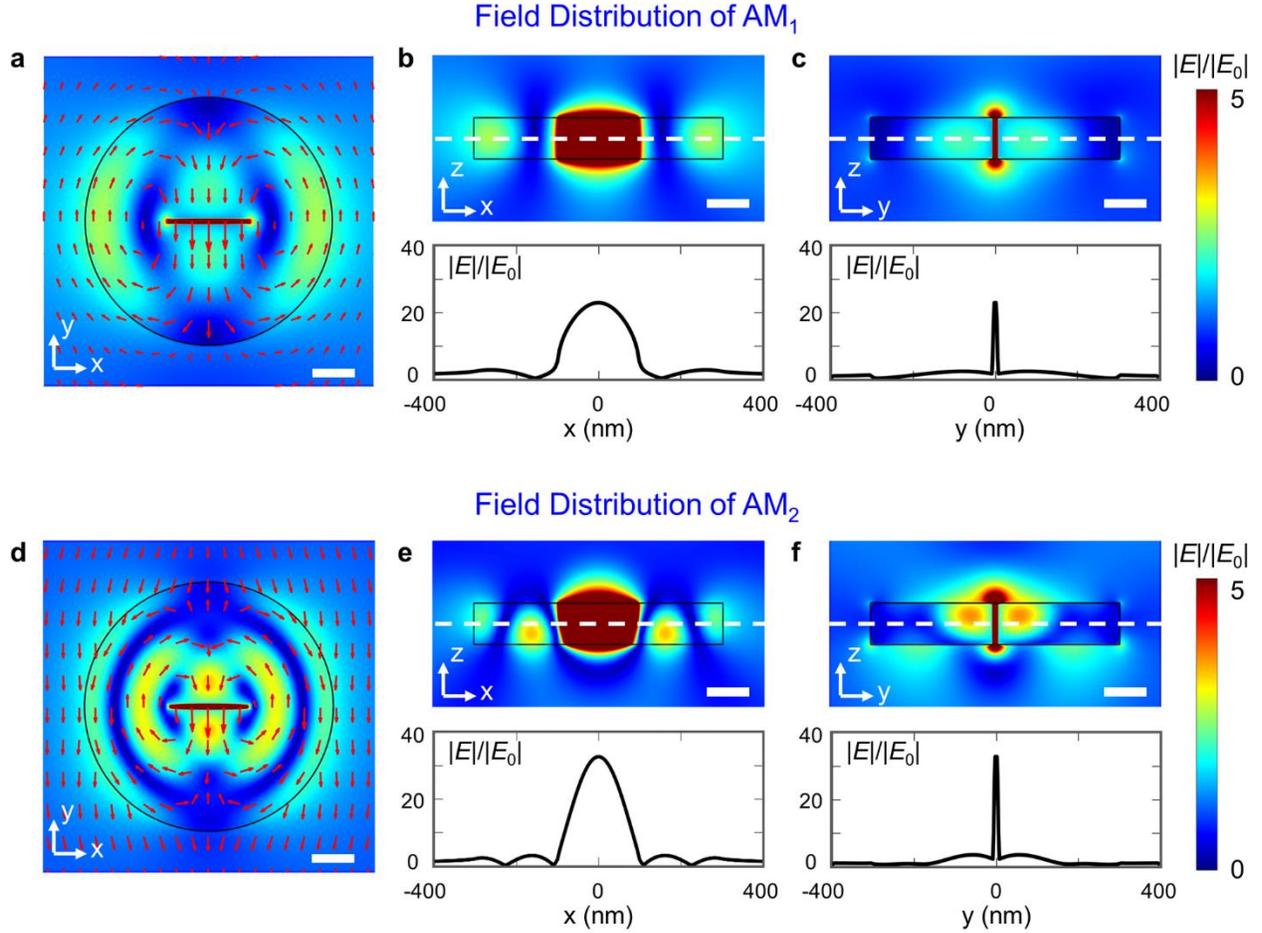

**Figure 3.** Electric field distributions at the $AM_1$ (a-c) and $AM_2$ (d-f) modes of the slotted disk, calculated at incident wavelength $\lambda$ = 1223 nm and 825 nm, respectively. To clearly show up the anapole features of the slotted disk, saturated fields (field magnitudes $|E|/|E_0| > 5$ have been set to 5) are provided intentionally. (a, d) Near field profiles on the $xy$ plane in the middle of the disk. (b, e) Near field profiles on the $xz$ plane through the center of the disk. The bottom curves show the unsaturated details of the enhanced fields along the $x$-axis (white dashed lines in the upper 2D plot). One can clearly see two and four field zeros along the $x$-cut line at the $AM_1$ and $AM_2$ modes, respectively. (e, f) Near field profiles on the $yz$ plane through the center of the disk. The bottom

curves show the unsaturated details and spatial confinement of the enhanced fields along the *y*-axis (white dashed lines in the upper 2D plot). The scale bars represent 100 nm in all panels.

Despite the similar nature of the anapole modes in both disks, the slotted disk exhibits a much brighter central hotspot with field enhancements ~ 25 at the $AM_1$ mode and ~ 35 at the $AM_2$ mode (Figure S4), whereas the solid disk only possesses enhancements ~ 3 and ~ 5 at the two anapoles, respectively (Figure S3). This dramatic field enhancement arising from the slot has a very simple physical interpretation: the boundary condition for the normal component of the electric displacement. Here in our case, the continuity of the *y* component of the electric displacement $D_y$ enforces the electric field $E_y$ inside the slot to feature a significantly higher amplitude than that on the Si side. As a result, one can clearly see the electric field abruptly "jumps" at the air-silicon interfaces along the *y*-axis, giving rise to an extreme confinement of the electric field within the slot volume (Figure 3c,f). Therefore, in addition to the anapole generation, the high refractive index of Si offers a simultaneously extra pathway for acquiring local field enhancement. This effect is also akin to that exploited in slow-plasmon resonant structures,[40] where a narrow (~ 5 nm) gap introduced at maxima of internal electric fields leads to a field enhancement $|E|/|E_0|$ ~ 30. Furthermore, our configuration provides an evenly distributed field enhancement inside the slot. The ratio between the maximum field enhancement $|E_{max}/E_0|$ and the average field enhancement $|\bar{E}_{slot}/E_0|$ inside the slot volume is 1.34:1 and 1.58:1 for the $AM_1$ and $AM_2$ modes, corresponding to an average intensity enhancement ~ 280 and ~ 400 folds, respectively. Compared to usual dielectric dimer gaps with sharp edges,[14-17] our design yields a much higher electric field enhancement along with a larger exposure area of the hotspots. This property affords a feasible way to realize a spatial overlap between hotspots and nearby emitters, which may find many applications in molecular sensing and quantum nanophotonics.[25]

The extra enhancement factor introduced by the boundary conditions is also influenced by the finite size of the slot, as shown in Figure 4. For length *L*, although the enhancement stays high against the variations, there is an optimized value for the highest field enhancement. This is because the existence of the slot slightly perturbs the near-field distributions of the conventional anapole modes. When the length is too small (e.g. $L$ = 50 nm), the slot cannot fully interact with the enhanced fields at the anapole modes and therefore cannot achieve a maximum field enhancement. Whereas if the length is too large (e.g. $L$ = 600 nm), the slot would interface the edges of the poloidal currents, where electric fields have reversed directions and thus are also unfavorable for maximizing the field enhancement. Only when the slot length is nearly equal to the distance between the two field zeros, such an impact would become minimized and a highest enhancement factor can thus be obtained. Therefore, the distinct field distributions at the $AM_1$ and $AM_2$ modes result in different optimized slot lengths. As shown in Figure 4b, the optimized slot length for the $AM_2$ mode ($L \approx D/3 = 200$ nm) is smaller than that for the $AM_1$ mode ($L \approx D/2 = 300$ nm) because the electric fields at the $AM_2$ mode are more confined in the center and the distance between the nearest field zeros is shorter than that of the $AM_1$ mode. It is also worth noting that, due to the standing wave character of the $AM_2$ mode (four field zeros along the $x$ axis, as in Figure 3e), the enhancement factor reaches a plateau around $L \approx 2D/3 = 400$ nm rather than directly decreases as for the $AM_1$ mode. Figure 4c shows the impacts of the slot width $W$. The slot width $W$ affects the field enhancement in the same manner as it usually does in plasmonic and dielectric gap antennas. A much higher (lower) field enhancement can be achieved by decreasing (increasing) the width of the slot (Figure 4d and Figure S5). For example, when the slot width $W$ is decreased from 10 nm to 4 nm, the intensity enhancement can correspondingly increase from 1000 to 2500 folds, which largely outperforms that of many plasmonic dimers with the same gap size.[16, 40-42]

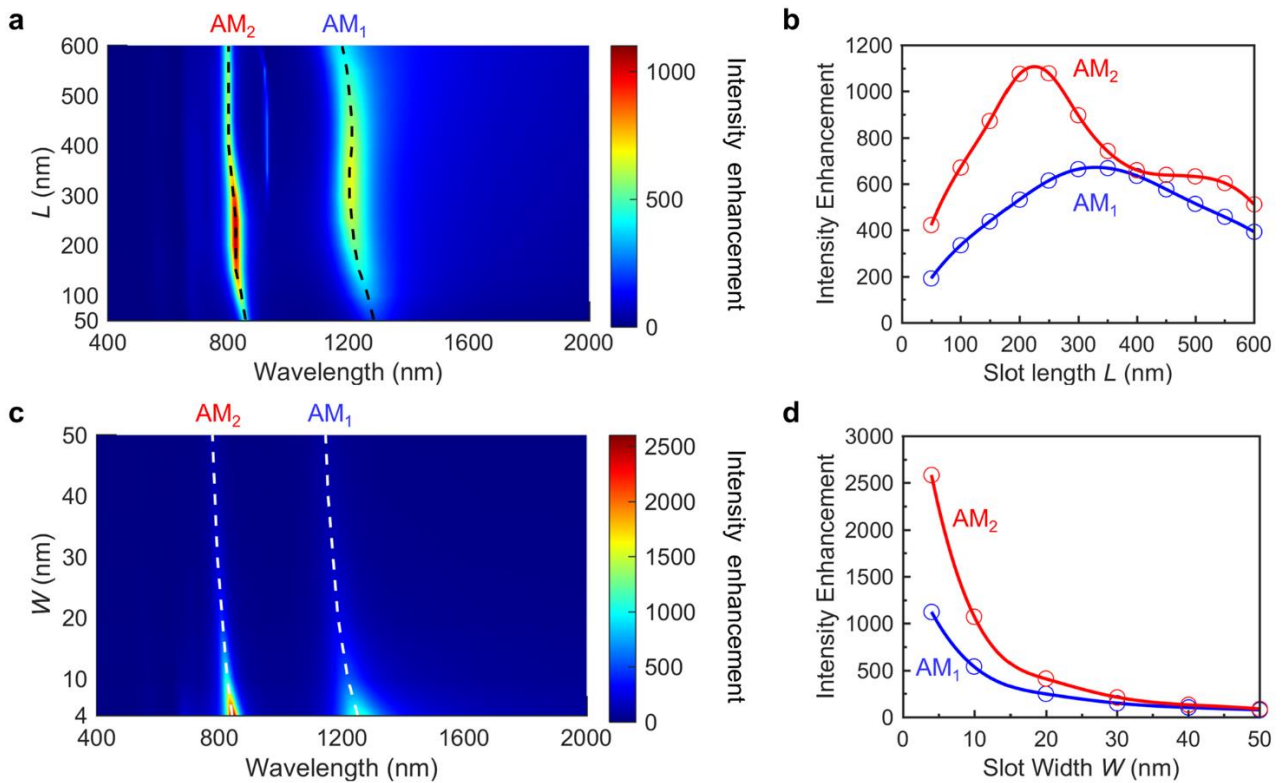

**Figure 4.** Impacts of the slot size on the electric intensity enhancement. The disk has the same size as in Figure 2. The slot length $L$ and width $W$ are fixed to 200 and 10 nm respectively when the other parameter is varied. (a,c) Two-dimensional plot of calculated intensity enhancements as functions of the wavelength and slot length $L$ (a) or width $W$ (c). The dashed lines indicate the peak enhancements at $AM_1$ and $AM_2$ modes, showing their overall blue-shift trends with increasing slot size. (b,d) Intensity enhancements at the $AM_1$ (blue circles) and $AM_2$ modes (red circles) with respect to the slot length $L$ (b) or the width $W$ (d).

Next, we investigate the impacts of the disk size on both far-field and near-field responses of the slotted disk. A continuous evolution of the spectra can be observed across the visible and near-infrared regions (Figure 5). As expected, the anapole modes exhibit a substantial red shift with increasing diameter $D$ or height $H$ in both far-field and near-field behaviors. One can also see that the field enhancements at shorter wavelengths are much smaller than those at longer wavelengths. One may simply attribute this phenomenon to the absorption loss of Si in the visible range. However, besides that, it is also worth noticing that the field enhancement continues increasing dramatically with increasing diameter $D$ in the near-infrared regime, where the absorption of silicon is negligible (Figure 5b). To better understand this interesting observation, we study the slotted disk with the largest diameter $D$ = 1000 nm in detail and implement multipole decomposition to examine its optical resonances. As shown in Figure S6, the largest disk indeed shows a much more significant intensity enhancement (~ 1800) compared to the disk with a smaller diameter $D$ = 600 nm (Figure 2), especially at the $AM_2$ mode. By further comparing their decomposed multipoles, one can see that, at the $AM_2$ modes of the largest disk ($\lambda$ ~ 1200 nm), there is only an inherently accompanied magnetic quadrupole existed with the anapole mode.[27, 28] Whereas, for the disk with a smaller diameter $D$ = 600 nm (Figure S1c), there is an extra magnetic dipole at the $AM_2$ mode ($\lambda$ ~ 800 nm), resulting in an additional channel for energy radiation and thus decreasing the field enhancement within the slot. Therefore, when $D$ increases, the anapole modes would become "purer" with fewer co-existed multipoles (i.e., more separated from other multipoles) and thereby give rise to higher intensity enhancements. This also explains why there is no such evident increase for the $AM_1$ mode, since the $AM_1$ mode is already "pure" enough for both disks with different diameters. We mention that the increase in diameter $D$ also gives rise to third- and fourth-order anapole modes, as $AM_3$ or $AM_4$ modes shown in Figure S6. Although the coexistence of other higher-order multipole terms decreases the stored electric energy, these higher-order anapole modes can still generate noticeable field enhancements. Further optimizations regarding the slot length *L* and width *W* can also be considered as discussed above to achieve higher and selective field enhancements in multi-spectral ranges.

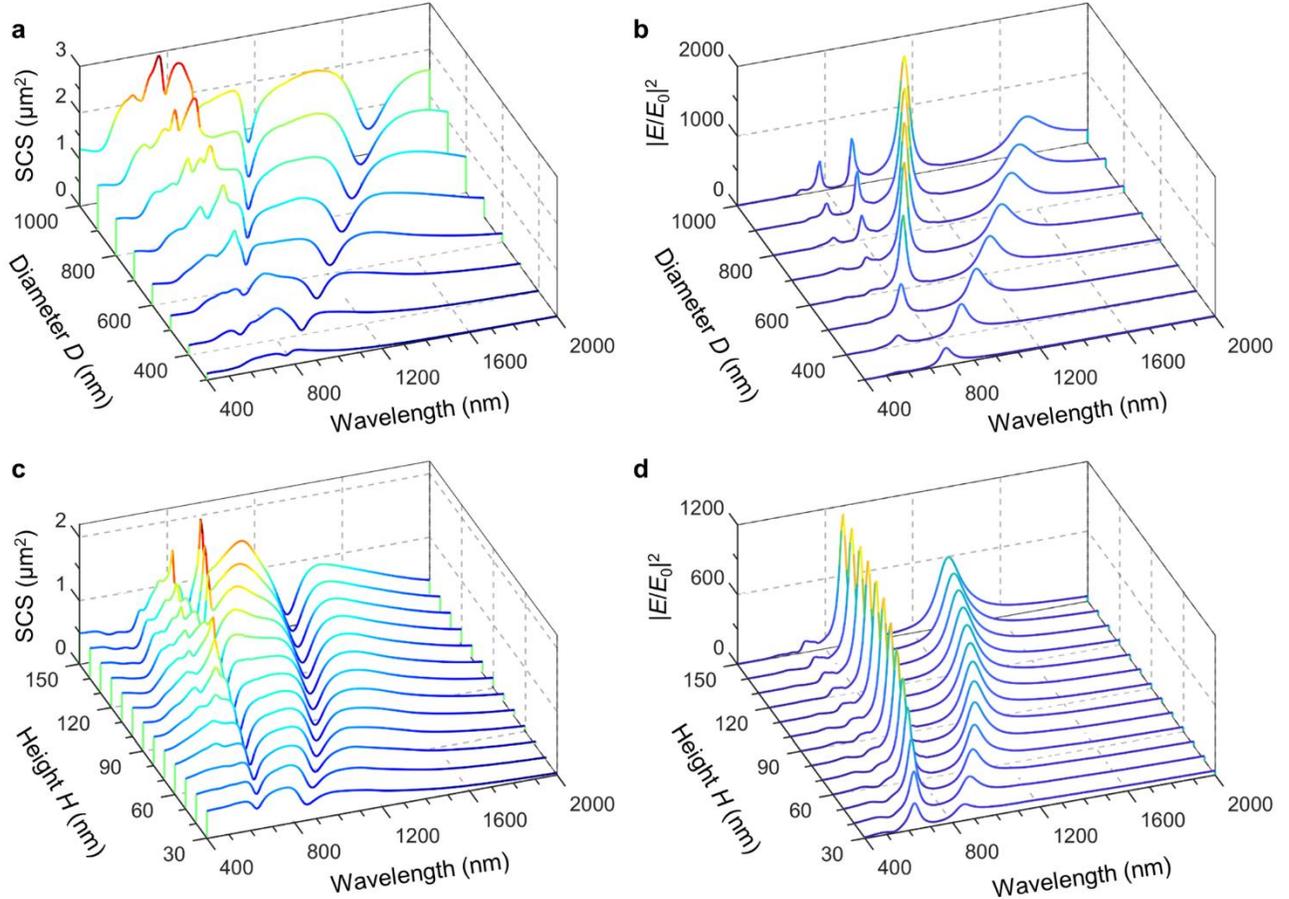

**Figure 5.** Size-dependent far-field scattering and near-field enhancement of the slotted disk. (a, c) Scattering spectra of a slotted disk with varying diameters *D* (a) or heights *H* (c). (b, d) Near-field intensity enhancement spectra of the slotted disk with varying diameters *D* (b) or heights *H* (d). For the panel (a) and (b), the height of the disk is fixed as $H = 100$ nm while for the panel (c) and (d), the diameter of the disk is fixed as $D = 600$ nm. In all above calculations, the width of the slot is fixed to $W = 10$ nm and the length *L* is set as $L = D/3$.

Likewise, if we decrease the height *H*, the nanodisk that has a higher diameter-to-height ratio would also support such "pure" anapole modes and possesses large field enhancements, even in the visible range given the non-negligible loss of Si (Figure 5d). To this end, we investigate the disk with a diameter $D = 600$ nm and a height $H = 30$ nm, as shown in Figure S7. Only a marginal magnetic dipole scattering can be observed with the $AM_2$ mode and the intrinsic magnetic quadrupole mode. This leads to a very high intensity enhancement (> 200 folds) in the visible region. When the gap size is decreased to 3 nm, an enhancement factor ~ 750 can be achieved. As a reference, optimized Si and Ag nanodimers with the same gap size (3 nm) only show intensity enhancements around 300 folds and 400 folds respectively at the same wavelength range.[15]

Besides the large field enhancement, another advantage of our design is the exposure of the anapole modes, making them directly accessible to nearby emitters. Here we consider spontaneous emission from an electric emitter characterized by a given dipole moment along the $y$-axis. The emitter has no intrinsic losses and is placed at the center of the slotted disk. As shown in Figure 6a, the radiative decay rate enhancement $\Gamma_R/\Gamma_0$ shows two clear peaks at different wavelengths, corresponding to the $AM_1$ and $AM_2$ modes. The produced rate enhancement (> 600) is much higher than a similar plasmonic nanodisk cavity.[43] Moreover, to maximize the emission enhancement, it is also essential to ensure the emitter spatially overlapped with the electric hotspots.[28] Given the relatively uniform field distribution within the slot, we can see that the radiative decay rate maintain high values inside the whole slot area (Figure 6b). In addition, the multi-resonant feature of our configuration also offers new possibilities to simultaneously enhance both excitation and emission processes at distinct wavelengths.

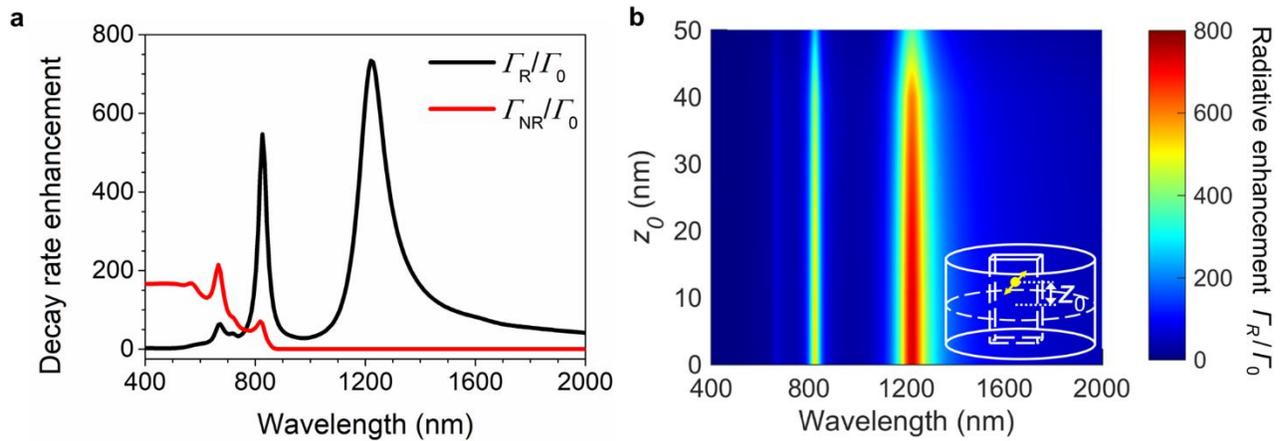

**Figure 6.** Slotted Si disk to enhance single photon emission. (a) Decay rate enhancement for a perfect dipole emitter in the center of the disk with the same dimensions as in Figure 2. Radiative (black) and non-radiative (red) decay rate enhancement are defined by the decay rate ($\Gamma_R$ or $\Gamma_{NR}$) normalized to the dipole's rate in free space $\Gamma_0$. (b) Two-dimensional map of the radiative decay rate enhancement as a function of wavelength and the relative position $z_0$ of the dipole along the $z$-axis. $z_0 = 0$ denotes the center of the disk.

**CONCLUSIONS**

In summary, we have demonstrated that strong electric fields with multi-resonant intensity enhancements exceeding $10^3$ times can be obtained within individual all-dielectric nanostructures such as a slotted Si nanodisk. By opening up a dedicated slot area at electric field maxima generated by anapole modes, electric field enhancement can be further boosted via boundary conditions without introducing any degradations in the far-field features of anapoles. Multipole decomposition method was used to thoroughly analyze both fundamental and higher-order optical responses of the system. The impacts of geometric parameters and mode interplays on spectral evolution and intensity enhancement were also systematically investigated. Our presented design methodology and associated effects are similar to those exploited in the plasmonic regime and can be readily extended to other high-index materials,[44, 45] other configurations with enhanced internal fields and other desired frequencies due to the scalable nature of Mie resonances.[46] In addition, we also showed that the direct accessibility of the hotspots leads to efficient enhancement of single-photon emission with large Purcell effects, bringing about entirely new applications of optical anapole modes. The uniform field distribution inside the proper-designed slot further makes it feasible to realize a large spatial overlap between quantum emitters and electric hotspots, which may find many applications in sensing and quantum nanophotonics.

**METHODS**

We performed three-dimensional FDTD simulations with a commercially software package (Lumerical). For all calculations, Si nanodisks was placed in a vacuum and perfectly matched layers were set as simulation boundaries to enclose the simulation area. The refractive index of the amorphous Si was determined by ellipsometry (Figure S1) and has been used in our previous study.[28] A mesh size of 5 nm was set over the whole volume of the Si disk and a finer override mesh of 1 nm was employed in the slot area. A normal-incident total-field/scattered-field planewave source was utilized to calculate the scattering cross section and intensity enhancement of the Si disks while an electric dipole source was exploited to calculate the radiative and nonradiative decay rate enhancement. To carry out multipole decomposition (see Supporting Information), a three-dimensional frequency-domain field monitor was used to record the electric fields inside the disks.

**Acknowledgements**

This work was funded by the European Research Council (the PLAQNAP project, Grant 341054) and the University of Southern Denmark (SDU2020 funding).

**Competing financial interests**

The authors declare no potential conflicts of interest.

**Supporting Information**

Multipole decomposition method; refractive index of amorphous silicon; comparison between slotted and solid disks; influence of the slot size; optical response of slotted disks with different geometric parameters.

# Supporting information

Anapole-Assisted Strong Field Enhancement in Individual All-Dielectric Nanostructures

*Yuanqing Yang,\* Vladimir A. Zenin, and Sergey I. Bozhevolnyi*

SDU Nano Optics, University of Southern Denmark, Campusvej 55, DK-5230 Odense, Denmark

\* Email address: yy@mci.sdu.dk.

## Multipole decomposition:

In order to rigorously analyze both fundamental and higher-order anapole modes excited in the slotted disks, here we employ spherical multipole decomposition, which can represent the total scattering cross section $C_{sca}$ of the particle as a sum of partial contribution from spherical multipoles[1]:

$$C_{sca} = \frac{\pi}{k^2} \sum_{l=1}^{\infty} \sum_{m=-l}^{l} (2l+1)(|a_E(l,m)|^2 + |a_M(l,m)|^2), \qquad (1)$$

where electric $a_E(l,m)$ and magnetic $a_M(l,m)$ spherical multipole coefficients can be directly calculated through the induced currents $\boldsymbol{J}(\boldsymbol{r}) = -i\omega\varepsilon_0[\varepsilon_r(\boldsymbol{r})-1]\boldsymbol{E}(\boldsymbol{r})$ as[1,2]:

$$a_E(l,m) = \frac{(-i)^{l-1}k\eta}{E_0\sqrt{\pi(2l+1)(l+1)}} \int Y_{lm}^*(\theta,\varphi) j_l(kr) \left\{ k^2 \boldsymbol{r}\cdot\boldsymbol{J}(\boldsymbol{r}) + \left(2+r\frac{d}{dr}\right)[\boldsymbol{\nabla}\cdot\boldsymbol{J}(\boldsymbol{r})] \right\} d^3r, \qquad (2)$$

$$a_M(l,m) = \frac{(-i)^{l-1}k^2\eta}{E_0\sqrt{\pi(2l+1)(l+1)}} \int Y_{lm}^*(\theta,\varphi) j_l(kr) \boldsymbol{r}\cdot[\boldsymbol{\nabla}\times\boldsymbol{J}(\boldsymbol{r})] d^3r, \qquad (3)$$

with $\omega_0$ and $E_0$ are the angular frequency and electric field amplitude of the incident plane wave, respectively; $\varepsilon_0$ is the vacuum permittivity; $\varepsilon_r(\boldsymbol{r})$ is the relative electric permittivity at any coordinate $\boldsymbol{r}$ with the disk's center at the origin point; $\boldsymbol{E}(\boldsymbol{r})$ is the electric field recorded in the 3D monitor, containing both incident field and scattered field; $k$ and $\eta$ are the wavenumber and the impedance of the surrounding medium, i.e., vacuum in our case; $Y_{lm}$ are the scalar spherical harmonics and $j_l(kr)$ are the spherical Bessel functions of first kind.

By implementing integration by parts and introducing the Riccati-Bessel functions $\Psi_l(kr)$ and the associated Legendre polynomials $P_l^m$ as defined in Jackson's textbook,[3] we can rewrite the equation (2) and (3) in the following forms without spatial derivatives:[2]

$$a_E(l,m) = \frac{(-i)^{l-1}k\eta}{2\pi E_0} \frac{\sqrt{(l-m)!}}{\sqrt{l(l+1)(l+m)!}} \int \exp(-im\phi) \left\{ [\Psi_l(kr)+\Psi_l''(kr)]P_l^m(\cos\theta)\hat{\boldsymbol{r}}\cdot\boldsymbol{J}(\boldsymbol{r}) \right.$$
$$\left. + \frac{\Psi_l'(kr)}{kr}\left[\frac{d}{d\theta}P_l^m(\cos\theta)\hat{\boldsymbol{\theta}}\cdot\boldsymbol{J}(\boldsymbol{r}) - \frac{im}{\sin\theta}P_l^m(\cos\theta)\hat{\boldsymbol{\phi}}\cdot\boldsymbol{J}(\boldsymbol{r})\right] \right\} d^3r, \qquad (4)$$

$$a_M(l,m) = \frac{(-i)^{l+1}k^2\eta}{2\pi E_0} \frac{\sqrt{(l-m)!}}{\sqrt{l(l+1)(l+m)!}} \int \exp(-im\phi) j_l(kr) \left[ \frac{im}{\sin\theta}P_l^m(\cos\theta)\hat{\boldsymbol{\theta}}\cdot\boldsymbol{J}(\boldsymbol{r}) \right.$$
$$\left. + \frac{d}{d\theta}P_l^m(\cos\theta)\hat{\boldsymbol{\phi}}\cdot\boldsymbol{J}(\boldsymbol{r}) \right] d^3r. \qquad (5)$$

We note that the above equations allow for calculating partial scattering from multipoles of arbitrarily high order. Here, by expanding the scattering response of the system rigorously up to $4^{th}$ order multipoles, excellent agreement between the simulation and multipole decomposition results can be obtained even at a very short wavelength ~ 400 nm (Figure S2c, d). As such, we can clearly identify the contributions from each multipole term and unambiguously reveal the origin of higher-order anapoles of the system.

**Supporting figures:**

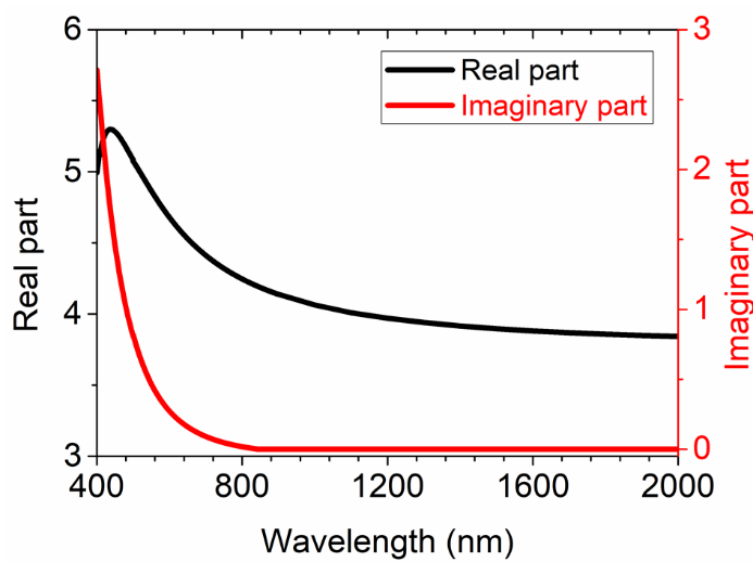

**Figure S1.** Refractive index of the amorphous silicon, produced by LPCVD growth and measured by ellipsometry.[4]

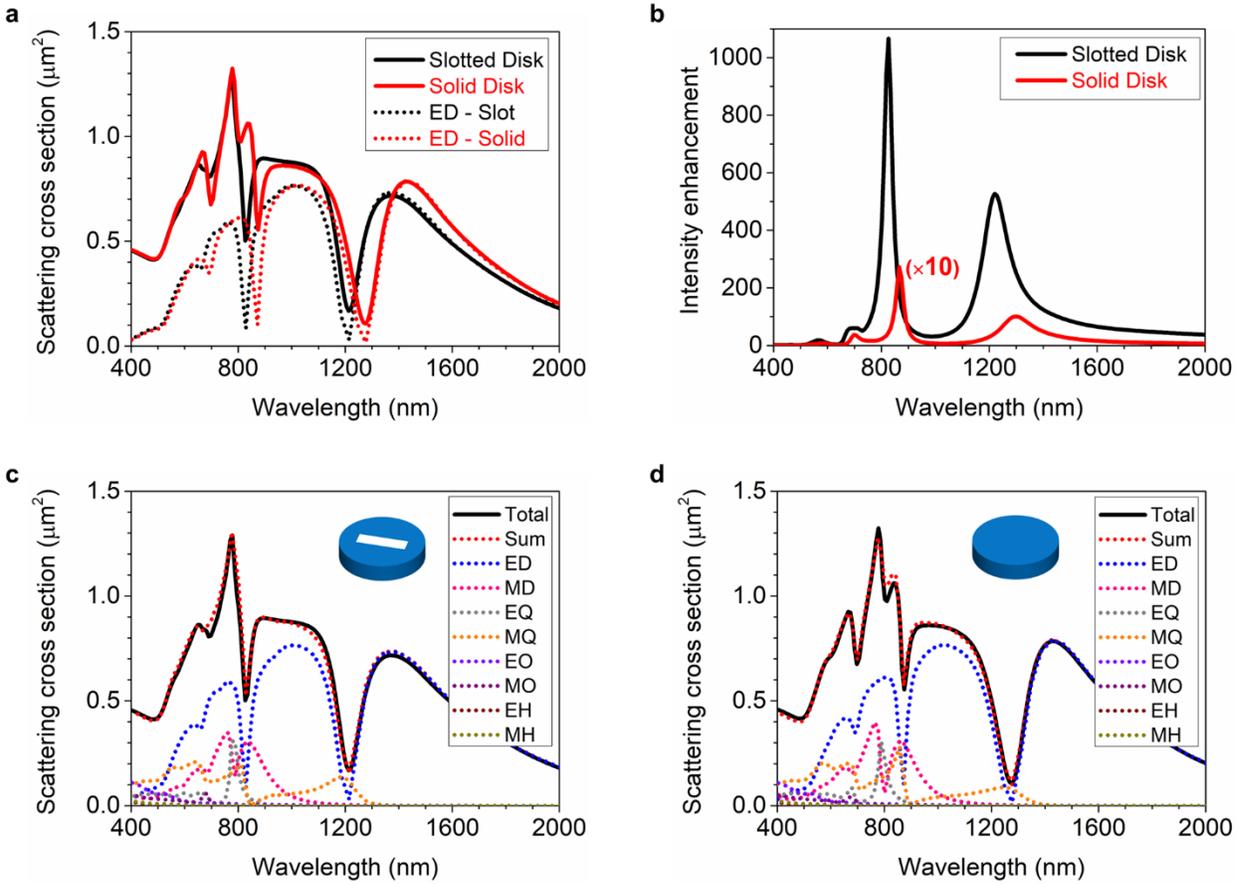

**Figure S2**. Comparison between the slotted and the solid nanodisks. The solid nanodisk has the same diameter ($D = 600$ nm) and the same height ($H = 100$ nm) as to the slotted disk in Figure 2. (a) Total scattering (solid lines) and partial contribution from the spherical electric dipole moment (dotted lines). Black lines represent the scattering response of the slotted disk while the red lines represent the scattering response of the solid disk. (b) Intensity enhancement of the electric field at the center of the slotted (black) and solid (red) disks. The response of the solid disk has been magnified by a factor of ten for better visibility. (c, d) Spherical multipole decomposition of the scattering cross sections of the slotted (c) and solid (d) disks, with simulated total scattering cross section ('Total', black lines), sum scattering from the decomposed multipoles up to the 4[th] order ('Sum', red dotted lines) and partial scattering from electric dipole (ED), magnetic dipole (MD), electric quadrupole (EQ), magnetic quadrupole (MQ), electric octupole (EO). magnetic octupole (MO), electric hexadecapole (EH) and magnetic hexadecapole (MH).

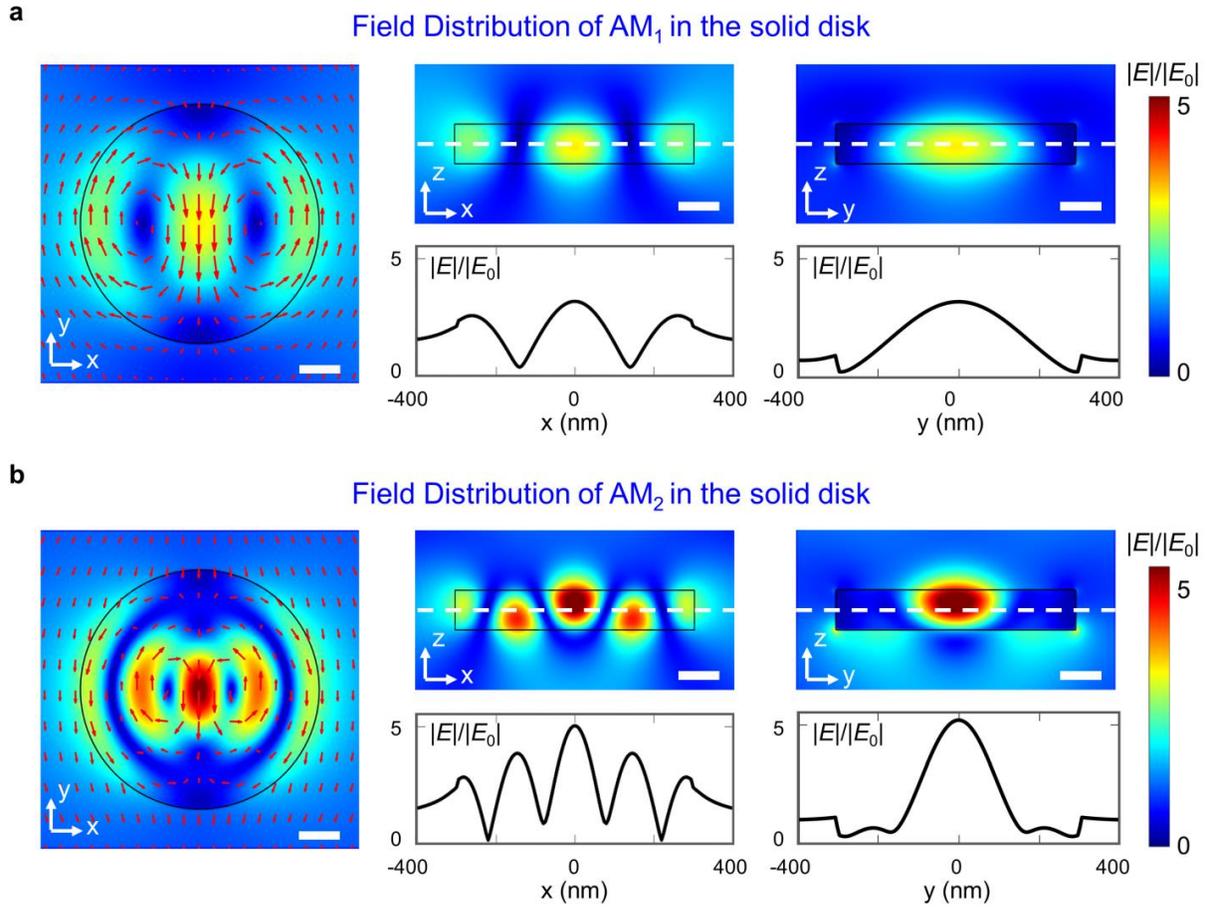

**Figure S3.** Electric field distributions of the AM$_1$ (a) and AM$_2$ (b) modes of the solid disk, calculated at the wavelength λ = 870 nm and 1310 nm, respectively. The color scale is the same as the Figure 3 in the main text. The scale bars represent 100 nm in all panels.

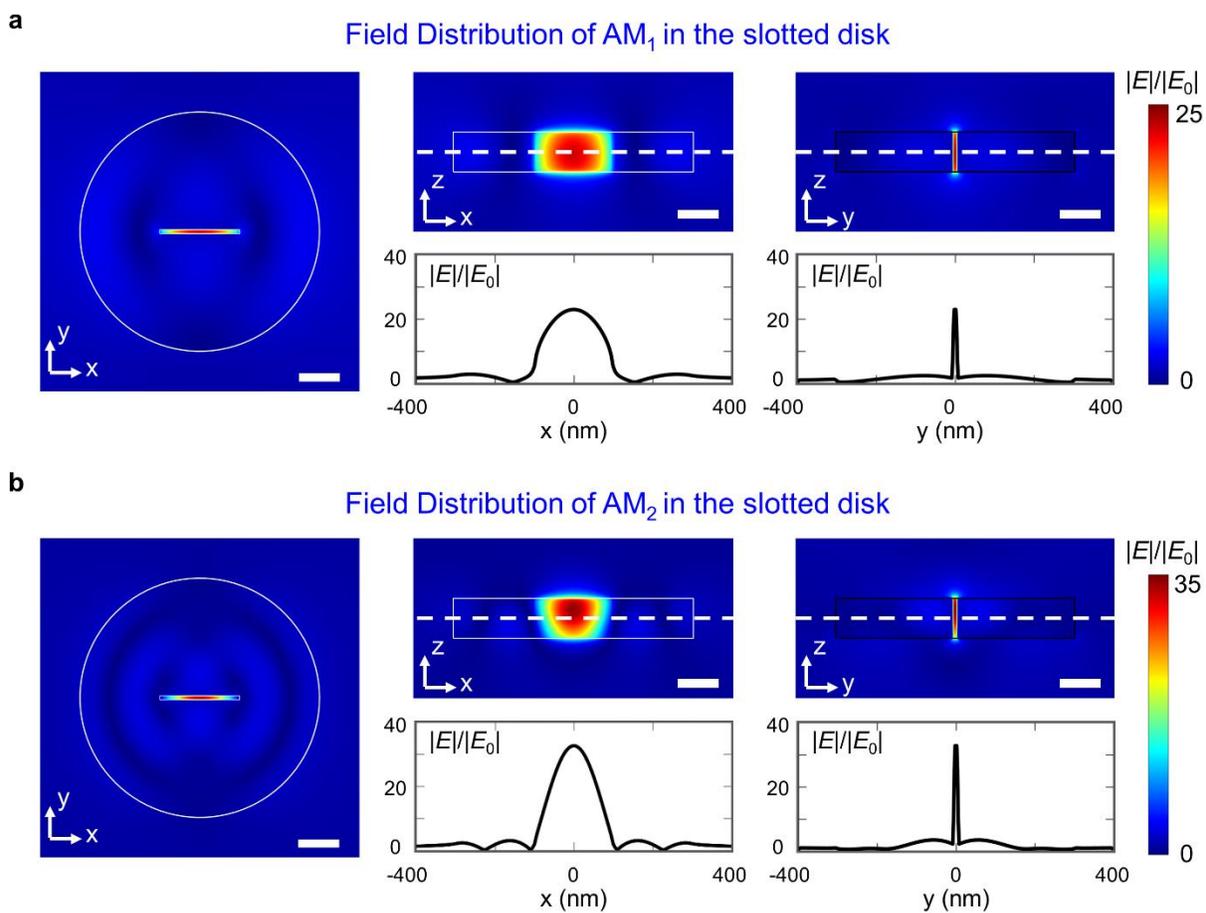

**Figure S4.** Unsaturated electric field distributions of the $AM_1$ (a) and $AM_2$ (b) modes of the slotted disk in Figure 3.

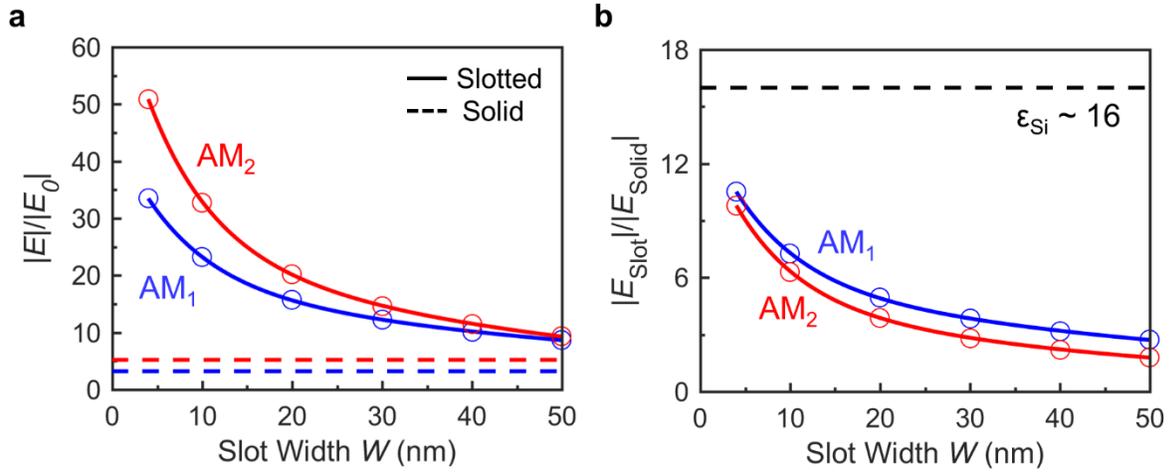

**Figure S5.** Field enhancement of the slotted disk as a function of the slot width *W*, with a reference of the solid disk. The disks have the same sizes as in Figure 4. (a) Total field enhancement $|E|/|E_0|$. Solid lines are for the slotted disk while the dashed lines indicate the field enhancements of solid disks at the two anapole modes. (b) Extra field enhancement $|E_{\text{slot}}|/|E_{\text{solid}}|$ introduced by the slot. The extra enhancement factor increases exponentially with decreasing slot width *W*, continuously approaching the theoretical limit predicted by the boundary conditions $\varepsilon_{Si}$ (~ 16 in the near-infrared region, see Figure S1).

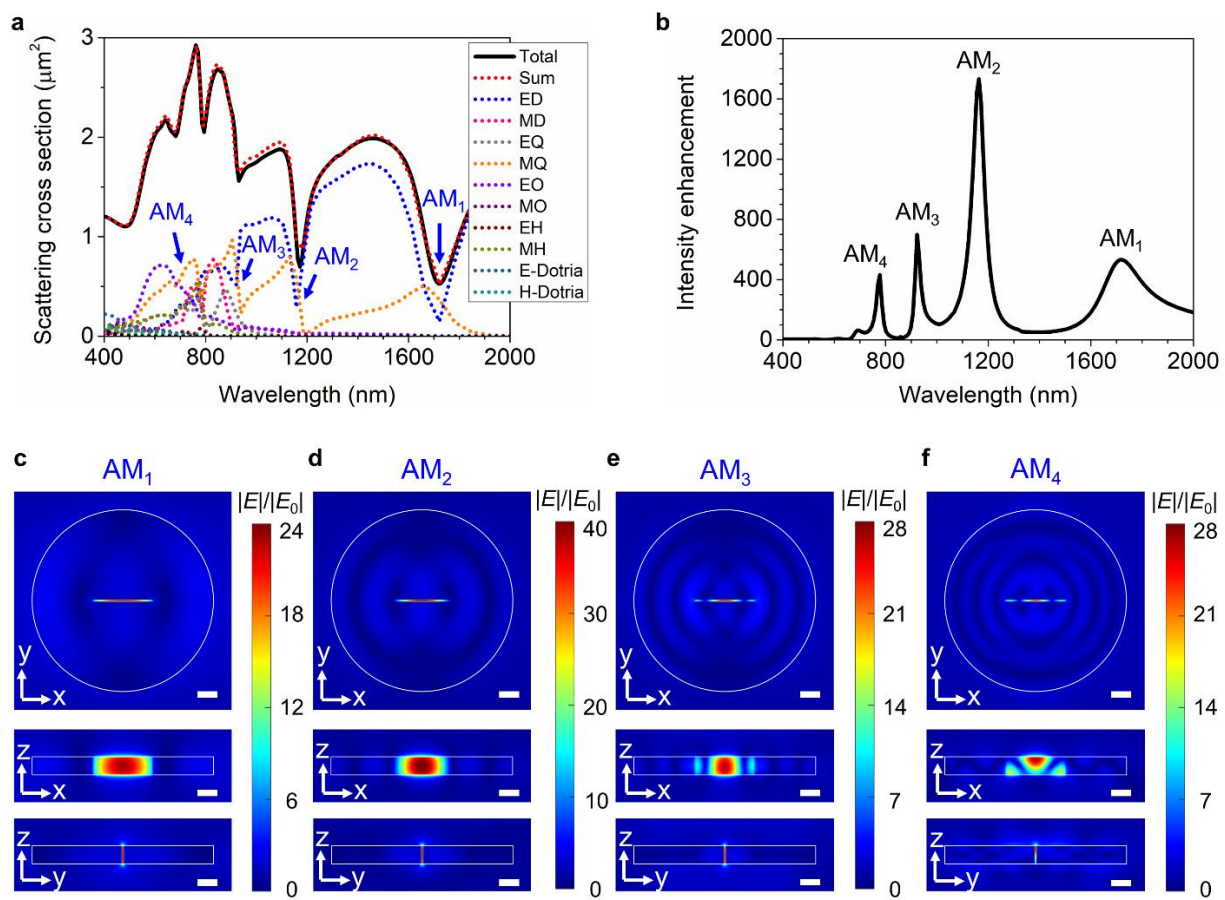

**Figure S6**. Far-field and near-field response of the largest slotted disks with $D = 1000$ nm, $H = 100$ nm, $L = 333$ nm and $W = 10$ nm. (a) Scattering cross section and its multipole decomposition of the slotted Si disk. E-Dotria and H-Dotria indicate the electric and magnetic dotriacontapole (32-poles) terms, respectively. (b) Multiresonant intensity enhancement at the center of the disk. (c-f) Electric field distributions of the four anapole modes supported by the disk. The scale bars in all panels indicate 100 nm.

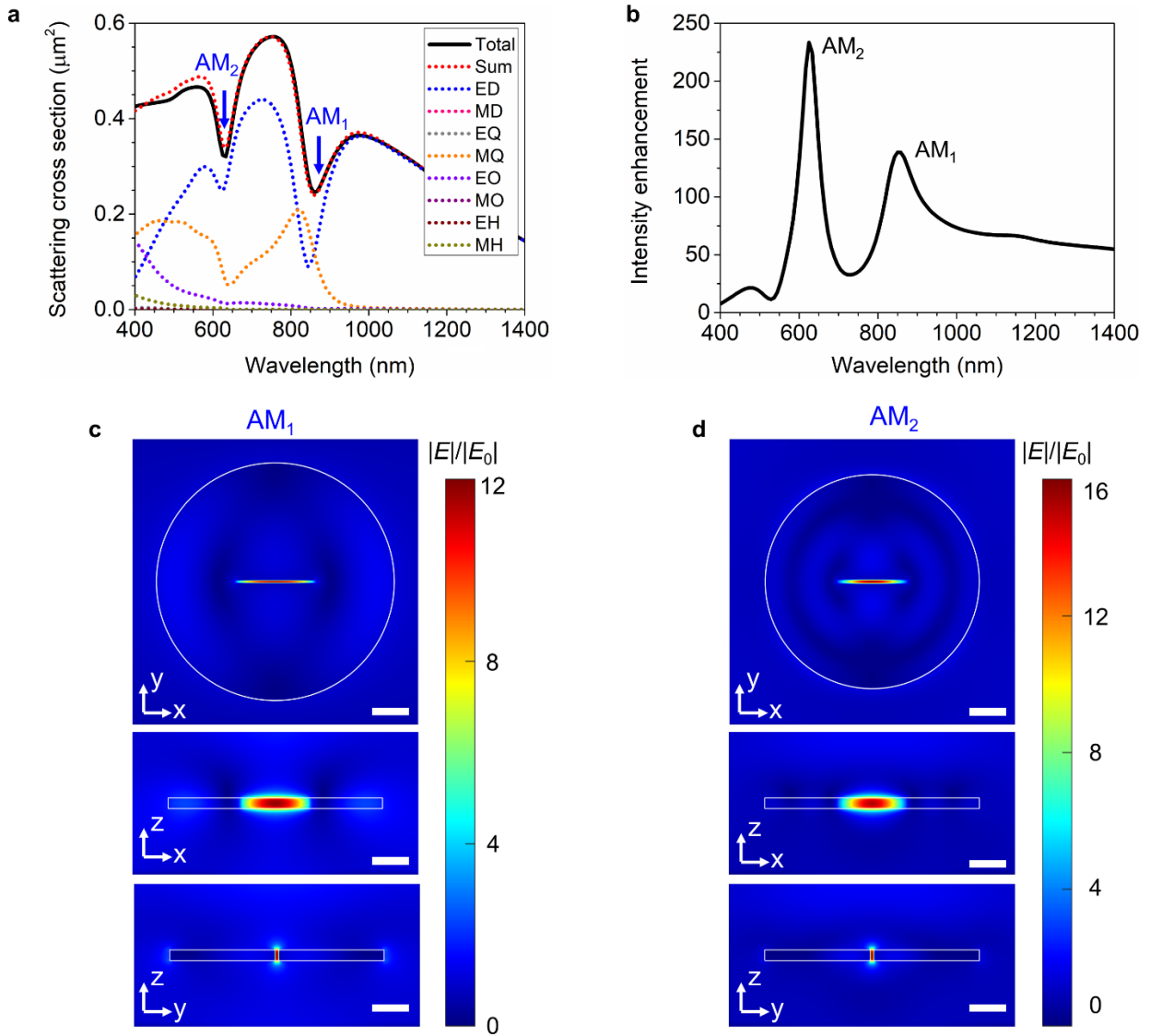

**Figure S7**. Far-field and near-field response of the thinnest slotted disks with $D = 600$ nm, $H = 30$ nm, $L = 200$ nm and $W = 10$ nm. (a) Scattering cross section and its multipole decomposition of the slotted Si disk. (b) Multiresonant intensity enhancement at the center of the disk. (c-f) Electric field distributions of the two anapole modes supported by the disk. The scale bars in all panels indicate 100 nm.